\documentclass[a4paper,amsmath,amssymb]{jpconf}
\usepackage{graphicx,color}
\usepackage{amsmath, amsthm, amssymb}
\begin{document}
\title{Dynamics of a flexible polymer in planar mixed flow}

\author{Dipjyoti Das$^1$, Sanjib Sabhapandit$^2$  and Dibyendu Das$^1$ }

\address{$^1$ Department of Physics, Indian Institute of Technology Bombay, Powai, Mumbai - 400076, India
}
\address{$^2$ Raman Research Institute, Bangalore 560080, India}

\ead{dipjyoti@phy.iitb.ac.in, sanjib@rri.res.in, dibyendu@phy.iitb.ac.in}

\begin{abstract}
We present exact spatio-temporal correlation functions of a Rouse
polymer chain submerged in a fluid having planar mixed flow, in the
steady state. Using these correlators, determination of the time scale
distribution functions associated with the first-passage tumbling
events is difficult in general; it was done recently in {\it
  Phys. Rev. Lett.} {\bf 101}, 188301 (2008), for the special case of
``simple shear'' flow. We show here that the method used in latter
paper fails for the general mixed flow problem. We also give many new
estimates of the exponent $\theta$ associated with the exponential
tail of the angular tumbling time distribution in the case of simple
shear.
\end{abstract}
\section{Introduction}
\label{sec:Introduction}

 Advancement in the techniques of fluorescence microscopy have led to
 direct observation and experimentation of single polymer chains (like
 a $\lambda$- DNA molecule) in fluid flows in the recent years
 \cite{chu98,chu99,writz_nat,tethered_prl,chu_angtumb1,chu_angtumb2,gera1,gera2}. Several
 works on theory and simulations have also addressed this problem,
 mostly for the case of ``simple shear'' flow
 \cite{degennes,tur1,tur2,tur3,winkler,delgado,das}. This problem has
 biological relevance \cite{bioappl_chu} and in particular plays a
 crucial role in the blood clotting process \cite{bioappl_vwf}. But
 planar fluid flows need not be simple shear only, and more
 complicated flows which are either predominantly elongational or rotational
  have also been studied \cite{degennes,chu_mixed,arti_mixed,mixed}.

\begin{figure}[ht]
\begin{center}
\includegraphics[width=11.9cm]{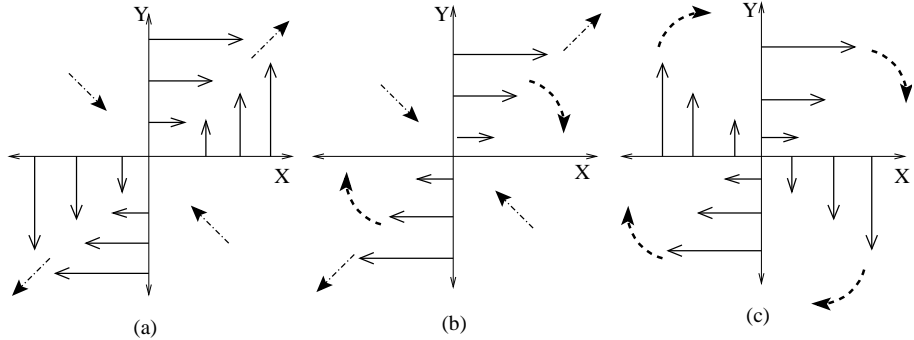}

\end{center}
\caption{\footnotesize 
 Schematic diagrams of different regimes of planar mixed flow. (a) The pure elongational flow $(\alpha=+1)$. (b) Simple shear $(\alpha=0)$. (c) The pure rotational flow $(\alpha =-1)$. Solid arrows represent the force vector field and dotted arrows represent the flow directions.}
\label{fig:fg1}
\end{figure}

In general, any planar flow of the form
$\mathbf{v}=v_x\hat{x}+v_y\hat{y}$ can be considered as a linear
superposition of a rotational component with vorticity $\omega =
\frac{1}{2}(\frac{\partial v_y}{\partial x}-\frac{\partial
  v_x}{\partial y} )$ and an elongational component with strain rate
$\dot{\epsilon}=\frac{1}{2}(\frac{\partial v_y}{\partial
  x}+\frac{\partial v_x}{\partial y} )$ \cite{degennes}. The ``mixed
flow'' refers to an intermediate flow between pure elongational flow
$(\omega=0)$ which creates large chain deformation, and pure
rotational flow $(\dot{\epsilon}=0)$ which is responsible for angular
rotation. ``Simple shear flow'' is the limit where the magnitude of
rotational component exactly equals that of elongatinal component i.e
$|\omega|=|\dot{\epsilon}|$. In this type of flow a polymer tumbles as
well as undergoes coil-stretch deformation. In Cartesian coordinates
the force vector in a mixed flow can be represented as:
$\mathbf{f}(t)\equiv (\dot{\gamma}y(t),\alpha \dot{\gamma} x(t),0)$
where $\dot{\gamma}$ is a constant rate. The parameter
$\alpha\in[-1,1]$ controls the relative magnitude of elongational and
rotational components. In the limit $\alpha \rightarrow 0$ this flow
reduces to simple shear flow in the $X$ direction. On the other hand,
the two extreme limits $\alpha \rightarrow +1$ and $\alpha \rightarrow
-1$ correspond to pure elongational flow and pure rotational flow
respectively (see Fig. \ref{fig:fg1}).

A polymer in a flow field has been modeled theoretically as Rouse
chain \cite{rouse,doi} with FENE end-to-end constraint
\cite{arti_mixed}, as a FENE dumbbell \cite{tur1,tur2} and as a liner
(but non-Rouse) chain with global contour length
constraint\cite{winkler}. Very recently we have shown \cite{das} that
analytically exact static distribution functions for the polar
coordinates of the polymer's end-to-end vector $\mathbf R(t)$ are
possible to be derived for a Rouse chain (without any nonlinear
constraint) in simple shear $(\alpha=0)$. Furthermore for the latter
flexible polymer, for the first time, we presented an analytical
calculation for the angular tumbling time distribution function. This
is very nontrivial as $\mathbf R(t)$ (although Gaussian) is a
non-Markovian process, for which in general calculation of
first-passage properties like tumbling are very difficult. It was
shown that the probability density function (PDF) of angular tumbling
time $\tau$, goes as $\exp (-\theta \tau/\tau_0)$ (where $\tau_0$ is
the polymer's longest relaxation time in the absence of any external
flow) and $\theta$ in general depends on the shear rate. In the limit
of strong shear the exponent $\theta$ goes to a constant value and it
was calculated (in Ref. \cite{das}) by a method known as ``independent
interval approximation'' (IIA)
\cite{mcfadden,majumder1,majumder2,majumder3}. The method of IIA was
applicable in the limit of strong shear as the $X$-component of
$\mathbf R(t)$ became a {\it smooth} process (defined later) in that
limit.

In this paper we study the problem of a linear Rouse polymer in
general mixed flow. We have derived the exact autocorrelation
functions for the Cartesian components of $\mathbf R(t)$. We show that
the $X$-component of $\mathbf R(t)$ is a {\it non-smooth} process and
hence analytically IIA method is not applicable to find the exponent
$\theta$ in the general mixed flow. In the special case of ``simple
shear'', we have also investigated this exponent in more detail (than
earlier \cite{das}) and got few comparative estimates of constant
$\theta$ using various available IIA schemes. For the general mixed
flow we have numerical estimates of $\theta$ from observed exponential
tail in the tumbling time distribution in our simulation.

\section{General correlator for polymer in mixed flow}

 We study the linear Rouse model \cite{rouse,doi} of a polymer chain
 of $N$ beads connected by harmonic springs (each has spring constant
 $k$) in a planar mixed flow. The position vector of $n$th bead
 $(n=1,2,...,N)$ at time $t$ is denoted by $\mathbf{r}_n(t)$. For
 internal beads $ (n=2,3,...,N-1)$ the position vector evolves via the
 Langevin equation :
\begin{eqnarray}
\frac{d\mathbf{r}_n(t)}{dt}=
-k(2\mathbf{r}_n-\mathbf{r}_{n+1}-\mathbf{r}_{n-1})+\mathbf{f}_n(t)+\boldsymbol\eta(n,t)
\label{eq:langevin1}
\end{eqnarray}
where the vector $\mathbf{f}_n(t) \equiv (\dot{\gamma}y_n(t),\alpha
\dot{\gamma} x_n(t),0)^T$ denotes the planar mixed flow force field
with the constant rate $\dot{\gamma}$. The vector
$\boldsymbol\eta(n,t)\equiv (\eta_1(n,t),\eta_2(n,t),\eta_3(n,t))^T$
is the thermal white noise with zero mean and a correlator
$\langle\eta_i(n,t)\eta_j(n',t')\rangle= \zeta
\delta_{ij}\delta_{n,n'}\delta(t-t')$, where $i,j=1,2,3$ and
$n=1,2,...,N$. The noise strength $\zeta$ is proportional to
the temperature and in the Eq. (\ref{eq:langevin1}) the force strengths
are scaled by viscosity. It is convenient to introduce dimensionless
so-called Weissenberg number $\text{Wi}=\dot{\gamma}\tau_0$, where
$\tau_0=N^2/k\pi^2$ is the longest relaxation time of the polymer in
the absence of any external flow. Note that the two end-beads (for
$n=1$ and $n=N$) feel only one-sided interaction of the harmonic
spring. The position vectors of these two beads can also be included
into Eq. $(1)$ if we add two hypothetical beads at $n=0$ and $n=N+1$
with the ``free boundary conditions'': 
$\mathbf{r}_0(t)=\mathbf{r}_1(t)$ and
$\mathbf{r}_{N+1}(t)=\mathbf{r}_N(t)$.

 In the large $N$ limit, the discrete index $n$ can be treated as a continuous variable $u$ and Eq. (\ref{eq:langevin1}) then becomes 
\begin{eqnarray}
\frac{\partial \mathbf{r}(u,t)}{\partial t}= k\frac{\partial^2
  \mathbf{r}(u,t)}{\partial u^2} +\mathbf{f}(u,t)+\boldsymbol\eta(u,t)
\label{eq:langevin2}
\end{eqnarray}
with the free boundary conditions : $\frac{\partial
  \mathbf{r}(u,t)}{\partial u}|_{u=0}= 0$ and $\frac{\partial
  \mathbf{r}(u,t)}{\partial u}|_{u=N}= 0$. In this continuum limit the
noise correlator and the force field are given by
$\langle\eta_i(u,t)\eta_j(u',t')\rangle= \zeta \delta_{ij}\delta
(u-u')\delta(t-t')$ and $\mathbf{f}(u,t) \equiv
(\dot{\gamma}y(u,t),\alpha \dot{\gamma} x(u,t),0)^T$ respectively. 

We solve Eq. (\ref{eq:langevin2}) by Fourier cosine decomposition
using the transformation :
\begin{eqnarray}
\tilde{\mathbf{r}}_p(t)= \frac{1}{N}\int_0^N  \cos \left( \frac{p \pi u}{N}\right) \mathbf{r}(u,t)~ du ~~~~~~~~~~~~~ p=0,1,2,...
\label{eq:cos-transform}
\end{eqnarray}
where each mode evolves independently. For the mode coordinates the
problem reduces to the single particle problem (i.e a single bead connected
to the origin by a harmonic spring in a planar mixed flow) with the modified
spring constant $k_p=p^2/\tau_0$ and noise strength $\zeta'=\zeta/2N$:
\begin{eqnarray}
\frac{\partial \tilde{\mathbf{r}}_p(t) }{\partial t}= -k_p\tilde{\mathbf{r}}_p(t)  +\tilde{\mathbf{f}}_p(t)+\tilde{\boldsymbol \eta}_p(t)
\label{eq:single-mode}
\end{eqnarray}
where $\tilde{\mathbf{f}}_p(t) \equiv
(\dot{\gamma}\tilde{y}_p(t),\alpha \dot{\gamma} \tilde{x}_p(t),0)^T$
and $\tilde{\boldsymbol \eta}_p(t)$ are transformed force vector and
white noise respectively. Now Eq. (\ref{eq:single-mode}) which are
coupled equations of the mode coordinates $\tilde{x}_p(t)$,
$\tilde{y}_p(t)$ and $\tilde{z}_p(t)$ can be solved using the method
of Laplace transform and we obtain the exact two-time correlators of
the mode components.  

Our main quantity of interest is the end-to-end vector defined as:
$\mathbf{R}(t) \equiv \mathbf{r}(N,t)-\mathbf{r}(0,t)$. Using the
inverse Fourier cosine transform we find that the correlators
associated with the Cartesian components of $\mathbf R(t)$ are related
to the correlators associated with the Cartesian mode components as
below (for finite time increment $\tau \geq 0$) :
\begin{eqnarray}
\langle  R_i(t) R_j(t+\tau)\rangle= 16 \sum_{p=1,3,5,...}^{\infty}\langle \tilde{ r}_{ip}(t)\tilde { r}_{jp}(t+\tau)\rangle
\label{eq:R-correl}
\end{eqnarray}
where $R_i$ and $\tilde{r}_{ip}$ are the Cartesian components of the
end-to-end vector and the $p$th mode vector respectively. From the
above equation various correlation functions associated with the
Cartesian components of $\mathbf R(t)$ can be obtained explicitly.

\subsection{Exact autocorrelation functions of the end-to-end vector}

Using Eq. (\ref{eq:R-correl}) we have derived the exact autocorrelation functions of the Cartesian components of $\mathbf{R}(t)$ in the stationary state limit $ (t \rightarrow \infty)$ with a finite time increment $\tau \geq 0$, both  in the $\alpha>0$ and $\alpha<0$ regimes separately.

\underline{Elongational regime ($\alpha>0$)}: In this case the elongational effect dominates over the rotational effect. We note an important point that in the elongational regime the asymptotic correlators of $\mathbf R(t)$ exist only under the condition: $p^2>\text{Wi}\sqrt{\alpha}$ (for $p=1,3,5,...$). Thus the correlators diverge for the lower modes and hence one would expect huge instability in the polymer's configuration; it will get infinitely stretched and will not attain a stable state. We give the results in the limit $t\rightarrow \infty$ below, and express them as functions of a scaled time-increment variable $T \equiv \tau / \tau_0$ :

\begin{eqnarray}
\langle R_x(t)R_x(t+\tau_0 T) \rangle  &=&  \frac{2\zeta\tau_0 }{N}\sum_{p=1,3,5,...}^{\infty} e^{-p^{2}T} \Bigl [ \Bigl\{\cosh(\text{Wi} \sqrt{\alpha} T) \left(\frac{p^2}{p^4-\text{Wi}^2\alpha}+\frac{1}{p^2}\right) \nonumber \\
&& +\frac{\sinh(\text{Wi} \sqrt{\alpha} T) \text{Wi}\sqrt{\alpha}}{p^4-\text{Wi}^2\alpha}\Bigr\} + \frac{1}{\alpha} \Bigl\{ \cosh(\text{Wi} \sqrt{\alpha} T) \left(\frac{p^2}{p^4-\text{Wi}^2\alpha}-\frac{1}{p^2}\right)\nonumber\\
&& +\frac{\sinh(\text{Wi} \sqrt{\alpha} T) \text{Wi}\sqrt{\alpha}}{p^4-\text{Wi}^2\alpha}\Bigr\}\Bigr ] 
\label{eq:correl1}
\end{eqnarray}

\begin{eqnarray}
\langle R_y(t)R_y(t+\tau_0 T) \rangle  &=&  \frac{2\zeta\tau_0 }{N}\sum_{p=1,3,5,...}^{\infty} e^{-p^{2}T}  \Bigl[ \Bigl\{\cosh(\text{Wi} \sqrt{\alpha} T) \left(\frac{p^2}{p^4-\text{Wi}^2\alpha}+\frac{1}{p^2}\right)\nonumber\\
&&+\frac{\sinh(\text{Wi} \sqrt{\alpha} T) \text{Wi}\sqrt{\alpha}}{p^4-\text{Wi}^2\alpha}\Bigr\} +\alpha \Bigl\{ \cosh(\text{Wi} \sqrt{\alpha} T) \left(\frac{p^2}{p^4-\text{Wi}^2\alpha}-\frac{1}{p^2}\right)\nonumber\\
&&+\frac{\sinh(\text{Wi} \sqrt{\alpha} T) \text{Wi}\sqrt{\alpha}}{p^4-\text{Wi}^2\alpha}\Bigr\} \Bigr]
\label{eq:correl2}
\end{eqnarray}

\begin{eqnarray}
\langle R_x(t)R_y(t+\tau_0 T) \rangle  &=&  \frac{2\zeta\tau_0 }{N}\sum_{p=1,3,5,...}^{\infty} e^{-p^{2}T}  \Bigl[ \sqrt{\alpha}\Bigr\{\sinh(\text{Wi} \sqrt{\alpha} T) \left(\frac{p^2}{p^4-\text{Wi}^2\alpha}+\frac{1}{p^2}\right)\nonumber\\
&&+\frac{\cosh(\text{Wi} \sqrt{\alpha} T) \text{Wi}\sqrt{\alpha}}{p^4-\text{Wi}^2\alpha}\Bigr\} +\frac{1}{\sqrt{\alpha}}\Bigr\{ \sinh(\text{Wi} \sqrt{\alpha} T) \left(\frac{p^2}{p^4-\text{Wi}^2\alpha}-\frac{1}{p^2}\right)\nonumber\\
&&+\frac{\cosh(\text{Wi} \sqrt{\alpha} T) \text{Wi}\sqrt{\alpha}}{p^4-\text{Wi}^2\alpha}\Bigr\} \Bigr]
\label{eq:correl3}
\end{eqnarray}

\begin{eqnarray}
\langle R_y(t)R_x(t+\tau_0 T) \rangle  &=&  \frac{2\zeta\tau_0 }{N}\sum_{p=1,3,5,...}^{\infty} e^{-p^{2}T}  \Bigl[ \sqrt{\alpha}\Bigl\{\sinh(\text{Wi} \sqrt{\alpha} T) \left(\frac{p^2}{p^4-\text{Wi}^2\alpha}-\frac{1}{p^2}\right)\nonumber\\
&&+\frac{\cosh(\text{Wi} \sqrt{\alpha} T) \text{Wi}\sqrt{\alpha}}{p^4-\text{Wi}^2\alpha}\Bigr\}+\frac{1}{\sqrt{\alpha}}\Bigl\{ \sinh(\text{Wi} \sqrt{\alpha} T) \left(\frac{p^2}{p^4-\text{Wi}^2\alpha}+\frac{1}{p^2}\right)\nonumber\\
&&+\frac{\cosh(\text{Wi} \sqrt{\alpha} T) \text{Wi}\sqrt{\alpha}}{p^4-\text{Wi}^2\alpha}\Bigr\} \Bigr] \label{eq:correl4}\\
\langle R_z(t)R_z(t+\tau_0T) \rangle  &=&  \frac{4\zeta\tau_0 }{N}\sum_{p=1,3,5,...}^{\infty}\frac{ e^{-p^{2}T}}{p^2} \label{eq:correl5}\\
\langle R_z(t)R_x(t+\tau_0 T) \rangle &=& \langle R_z(t)R_y(t+\tau_0 T) \rangle=0 \label{eq:correl6}
\end{eqnarray}

\underline{Rotational regime $(\alpha<0)$}: In this regime the amount of vorticity is more dominant than the elongational component. We find the correlators in the limit $t\rightarrow \infty$ to be the following :

\begin{eqnarray}
\langle R_x(t)R_x(t+\tau_0T) \rangle  &=&  \frac{2\zeta\tau_0 }{N}\sum_{p=1,3,5,...}^{\infty} e^{-p^{2}T}  \Bigl[ \Bigl\{\cos(\text{Wi} \sqrt{-\alpha} T) \left(\frac{p^2}{p^4-\text{Wi}^2\alpha}+\frac{1}{p^2}\right) \nonumber\\
&& - \frac{\sin(\text{Wi} \sqrt{-\alpha} T) \text{Wi}\sqrt{-\alpha}}{p^4-\text{Wi}^2\alpha}\Bigr\}+\frac{1}{\alpha}\Bigl\{ \cos(\text{Wi} \sqrt{-\alpha} T) \left(\frac{p^2}{p^4-\text{Wi}^2\alpha}-\frac{1}{p^2}\right)\nonumber\\
&& - \frac{\sin(\text{Wi} \sqrt{-\alpha} T) \text{Wi}\sqrt{-\alpha}}{p^4-\text{Wi}^2\alpha}\Bigr\} \Bigr]
\label{eq:correl1p}
\end{eqnarray}

\begin{eqnarray}
\langle R_y(t)R_y(t+\tau_0T) \rangle  &=&  \frac{2\zeta\tau_0 }{N}\sum_{p=1,3,5,...}^{\infty} e^{-p^{2}T}  \Bigl[ \Bigl\{\cos(\text{Wi} \sqrt{-\alpha} T) \left(\frac{p^2}{p^4-\text{Wi}^2\alpha}+\frac{1}{p^2}\right) \nonumber\\
&& - \frac{\sin(\text{Wi} \sqrt{-\alpha} T) \text{Wi}\sqrt{-\alpha}}{p^4-\text{Wi}^2\alpha}\Bigr\} +\alpha\Bigl\{ \cos(\text{Wi} \sqrt{-\alpha} T) \left(\frac{p^2}{p^4-\text{Wi}^2\alpha}-\frac{1}{p^2}\right) \nonumber\\
&&- \frac{\sin(\text{Wi} \sqrt{-\alpha} T) \text{Wi}\sqrt{-\alpha}}{p^4-\text{Wi}^2\alpha}\Bigr\} \Bigr]
\label{eq:correl2p}
\end{eqnarray}

\begin{eqnarray}
\langle R_x(t)R_y(t+\tau_0T) \rangle  &=&  \frac{2\zeta\tau_0 }{N}\sum_{p=1,3,5,...}^{\infty} e^{-p^{2}T}  \Bigl[-\sqrt{-\alpha}\Bigl\{\sin(\text{Wi} \sqrt{-\alpha} T) \left(\frac{ p^2}{p^4-\text{Wi}^2\alpha}+\frac{1}{p^2}\right)\nonumber\\
&&+\frac{\cos(\text{Wi} \sqrt{-\alpha} T) \text{Wi}\sqrt{-\alpha}}{p^4-\text{Wi}^2\alpha}\Bigr\} +\frac{1}{\sqrt{-\alpha}}\Bigl\{ \sin(\text{Wi} \sqrt{-\alpha} T) \left(\frac{p^2}{p^4-\text{Wi}^2\alpha}-\frac{1}{p^2}\right)\nonumber\\
&&+\frac{\cos(\text{Wi} \sqrt{-\alpha} T) \text{Wi}\sqrt{-\alpha}}{p^4-\text{Wi}^2\alpha}\Bigr\} \Bigr]
\label{eq:correl3p}
\end{eqnarray}

\begin{eqnarray}
\langle R_y(t)R_x(t+\tau_0T) \rangle  &=&  \frac{2\zeta\tau_0 }{N}\sum_{p=1,3,5,...}^{\infty} e^{-p^{2}T}  \Bigl[-\sqrt{-\alpha}\Bigl\{\sin(\text{Wi} \sqrt{-\alpha} T) \left(\frac{p^2}{p^4-\text{Wi}^2\alpha}-\frac{1}{p^2}\right)\nonumber\\
&&+\frac{\cos(\text{Wi} \sqrt{-\alpha} T) \text{Wi}\sqrt{-\alpha}}{p^4-\text{Wi}^2\alpha}\Bigr\} +\frac{1}{\sqrt{-\alpha}}\Bigl\{ \sin(\text{Wi} \sqrt{-\alpha} T) \left(\frac{p^2}{p^4-\text{Wi}^2\alpha}+\frac{1}{p^2}\right)\nonumber\\
&&+\frac{\cos(\text{Wi} \sqrt{-\alpha} T) \text{Wi}\sqrt{-\alpha}}{p^4-\text{Wi}^2\alpha}\Bigr\} \Bigr] \label{eq:correl4p}\\
\langle R_z(t)R_z(t+\tau_0T) \rangle  &=&  \frac{4\zeta\tau_0 }{N}\sum_{p=1,3,5,...}^{\infty}\frac{ e^{-p^{2}T}}{p^2} \label{eq:correl5p}\\
\langle R_z(t)R_x(t+\tau_0 T) \rangle &=& \langle R_z(t)R_y(t+\tau_0 T) \rangle = 0 \label{eq:correl6p}
\end{eqnarray}

Note that these correlators (Eqs. \ref{eq:correl1p} - \ref{eq:correl6p} ) are in fact the analytic continuation of the results in Eqs.  \ref{eq:correl1} - \ref{eq:correl6} for $\alpha >0$ to $\alpha <0$. We also note that from the above list of the correlation functions we can obtain any static correlator by setting $T=0$ in relevant equation. To investigate any dynamic properties (like tumbling) we need to keep $T\neq 0$.

\subsection{Static distributions}

Since the Cartesian components of $\mathbf{R}(t)=(R_x(t),R_y(t),R_z(t))^T$ are all Gaussian variables, their joint PDF is also Gaussian and given by: $\mathcal{P}(R_x,R_y,R_z)=(2\pi)^{-3/2}|\mathbf C|^{-1/2}\exp(-\frac{1}{2}\mathbf {R}^T \mathbf{C} \mathbf{R})$, where $\mathbf C$ denotes the covariance matrix involving the static correlators. We also can obtain the joint PDF of the polar components of $\mathbf R(t)$ using the standard transformation: $\widetilde{\mathcal{P}}(R,\vartheta,\phi)=\mathcal{P}(R \cos{\vartheta} \cos{\phi} , R \cos{\vartheta} \sin{\phi} , R \sin{\vartheta}) \times R^2 \cos{\vartheta}$. In particular the azimuthal angle $(\phi)$ distribution $Q(\phi)$ provides some physical insight. Integrating $\widetilde {\mathcal{P}}(R,\vartheta,\phi)$ over $R$ we get the joint angular PDF $S(\vartheta,\phi)$ and again integrating it over $\vartheta$ we finally get the $\phi$-distribution, $Q(\phi)$.

\underline{In the elongational regime $(\alpha>0)$}: The covariance matrix is:

\begin{equation}
\mathbf {C} =\left( \begin{array}{ccc}
\langle R_x^2 \rangle & \langle R_xR_y \rangle & \langle R_xR_z \rangle \\
\langle R_yR_x \rangle & \langle R_y^2 \rangle & \langle R_yR_z \rangle \\
\langle R_zR_x \rangle & \langle R_zR_x \rangle & \langle R_z^2 \rangle \end{array} \right)= \left( \begin{array}{ccc}
d & b & 0 \\
b & a & 0 \\
0 & 0 & c \end{array} \right)
\label{eq:elong-matrix}
\end{equation}
and the determinant is $|\mathbf {C}|= af$, where $f=ad-b^2$; for the exact expressions of $d,...,c$ see  \ref{sec:AppA}. Finally the azimuthal angle distribution is :
\begin{eqnarray}
Q(\phi)=\frac{|\mathbf C|}{2\pi c \sqrt{f}}\frac{1}{(a \cos^2 \phi+ d \sin^2 \phi - b \sin 2\phi)}
\label{eq:phi-dist}
\end{eqnarray}
The most probable value of azimuthal angle from the above equation is given by : $\phi_m=\frac{1}{2} \tan^{-1}\left(\frac{2 b}{d-a}\right)$. Now, in the pure elongational limit $\alpha \rightarrow +1$ we get $d \approx a$ (see  \ref{sec:AppA}) and therefore $\phi_m \rightarrow \pi /4$. So the polymer stays along the $\phi=\pi/4$ plane for most of the time in this case which is expected intuitively. We can calculate the full width at half maximum $\Delta \phi$ of $Q(\phi)$ near $\phi_m \sim \pi/4 $ and it is given by : $\cos (2\Delta \phi)=2-(a/b)$. For  $\text{Wi} \rightarrow \infty$, we get $a/b \sim 1$ and hence $\Delta \phi \rightarrow 0$: so  the $Q(\phi)$ is sharply localised which is quite expected since the pure elongational flow is vorticity free where the polymer hardly tumbles. But such a limit is not actually realisable as $\text{Wi}\rightarrow \infty$ violates the mode stability condition $p^2>\text{Wi}\sqrt{\alpha}$, and so such a aligned polymer will experience an immediate catastrophic stretching.

\underline{In the rotational regime $(\alpha<0)$}: In this case we write the covariance matrix as :
\begin{equation}
\mathbf {\bar{C}} = \left( \begin{array}{ccc}
\bar{d} & \bar{b} & 0 \\
\bar{b} & \bar{a} & 0 \\
0 & 0 & \bar{c} \end{array} \right)
\label{eq:rot-matrix}
\end{equation}
and again the explicit values of $\bar{a},\bar{b},\bar{c}$ and  $\bar{d}$ are given in  \ref{sec:AppA}. The azimuthal distribution is : $\bar{Q}(\phi)=\left(\frac{|\mathbf {\bar{C}}|}{2\pi \bar{c} \sqrt{\bar{f}}} \right)/(\bar{a} \cos^2 \phi+ \bar{d} \sin^2 \phi - \bar{b} \sin 2\phi)$. In pure rotational limit $\alpha \rightarrow -1$, we get $\bar{d}\approx \bar{a}$ and $\bar{b} \rightarrow 0$, hence we have $\bar{Q}(\phi)=1/2\pi=$constant. Clearly, the pure vorticity dominated limit is quite expected to have a flat $\phi$ distribution. Moreover as $\alpha \rightarrow -1$, $\bar{d} \approx \bar{a} \approx \bar{c}$ and $\bar{b}\rightarrow 0$, leading to all eigenvalues of the matrix $\mathbf{\bar{C}}$ becoming identical. This means that the polymer approaches a spherical shape as $\alpha \rightarrow -1$ (pure rotational flow).

\subsection{Dynamic properties}
Now we concentrate on the first-passage question of the ``tumbling'' process of a polymer in mixed flow. The tumbling event is either defined as the successive occurrence of coiled states (radial tumbling) \cite{gera1,tur1} or as successive crossing of a fixed plane, say $\phi=0$, by the vector $\mathbf R(t)$ (angular tumbling) \cite{chu_angtumb1,chu_angtumb2,tur1}. The definition of radial tumbling is subjective, since a tumbling event is marked by $|\mathbf R(t)|$ falling below a value $R_0$ that defines the coiled state. But the angular tumbling time $\tau$ is objective and is simply the interval between two successive zero crossings of the process $R_x(t)$. The PDF of $\tau$ asymptotically goes as $P(T)\sim \exp (-\theta T)$, where $T\equiv \tau/\tau_0$ is the scaled time-increment.

The mean density of zero crossings for a Gaussian stationary process is given by: $\rho =1/\langle \tau \rangle = \sqrt{-A''(0)}/\pi$, where $A(\tau)$ is the normalised correlator of that process and $\tau$ denotes the intervals between zeros. The quantity $\rho$ is finite when $A'(0)=0$ and $A''(0)$ is finite. Such processes are called ``smooth'' and for them one can apply IIA to find the nontrivial exponent of the asymptotic distribution of the intervals between zeros \cite{mcfadden,majumder1}.

In the present case we are only interested in angular tumbling of the polymer. In the elongational regime ($\alpha>0$) angular tumbling is very rare event in the limit $\alpha \rightarrow +1$ due to sharp localisation of $\phi$-distribution for which $\phi$ hardly changes sign. To investigate the angular tumbling for $\alpha<0$ we proceed with the autocorrelation function  $C_{xx}(T)=\lim_{t \to \infty}\langle R_x(t)R_x(t+\tau_0 T) \rangle$ (see subsection $2.1$) and consider the normalised correlator $A(T)=C_{xx}(T)/C_{xx}(0)$. To know the small $T$ behavior of $A(T)$ we first obtain the exact Laplace transform of $C_{xx}(T)$ :

\begin{eqnarray}
\widetilde{C}_{xx}(s)  &=& \frac{2\zeta\tau_0}{N} \bigl[ \frac{1}{s}\frac{\pi}{8\sqrt{\beta}} \left( \tanh \frac{\pi\sqrt{\beta}}{2}+\tan \frac{\pi\sqrt{\beta}}{2}\right)\left(1+\frac{1}{\alpha}\right)+\frac{\pi^2}{8}\left(\frac{s}{s^2-\beta^2}\right)\left(1-\frac{1}{\alpha}\right)  \nonumber\\
 && ~~~~~~~~~~~~~  -\frac{\pi}{8\sqrt{s+\beta}}\left( \frac{2}{s+\beta}+\frac{\beta}{s(s+\beta)}(1+1/\alpha) \right)\tanh \left(\frac{\pi\sqrt{s+\beta}}{2} \right)\nonumber\\
  && ~~~~~~~~~~~~~  -\frac{\pi}{8\sqrt{s-\beta}}\left( \frac{2}{s-\beta}-\frac{\beta}{s(s-\beta)}(1+1/\alpha) \right)\tanh \left(\frac{\pi\sqrt{s-\beta}}{2}\right) \bigr] 
\label{eq:laplace}
\end{eqnarray}
where $\beta=\text{Wi}\sqrt{\alpha}$. From the above equation, the expression of $\tilde{A}(s) $ in large $s$ limit (i.e $s \rightarrow \infty$) can be obtained easily and inverting that we derive the expression of A(T) for $T\rightarrow 0$ as below:
 \begin{eqnarray}
A(T) &=&  1 -\frac{2\zeta\tau_0}{N C_{xx}(0)}\sqrt{\pi}  T^{1/2} - \frac{2\zeta\tau_0}{N C_{xx}(0)}\frac{\text{Wi}^2(-\alpha) \pi^2}{16}\left(1-\frac{1}{\alpha}\right) T^2+ \dotsb
\label{eq:smallT-mixed}
\end{eqnarray}
where $C_{xx}(0)=\langle R_x^2\rangle$ is the static correlator given
in \ref{sec:AppA}.
It is clear from the above equation that both $A'(0)$ and $A''(0)$ diverge. Thus $R_x(t)$ is a ``non-smooth'' process in the sense that the density of zero crossings is infinite and hence the method of IIA can not be applied. 

\begin{figure}[ht]
\begin{center}
\includegraphics[width=10.5cm,angle=0]{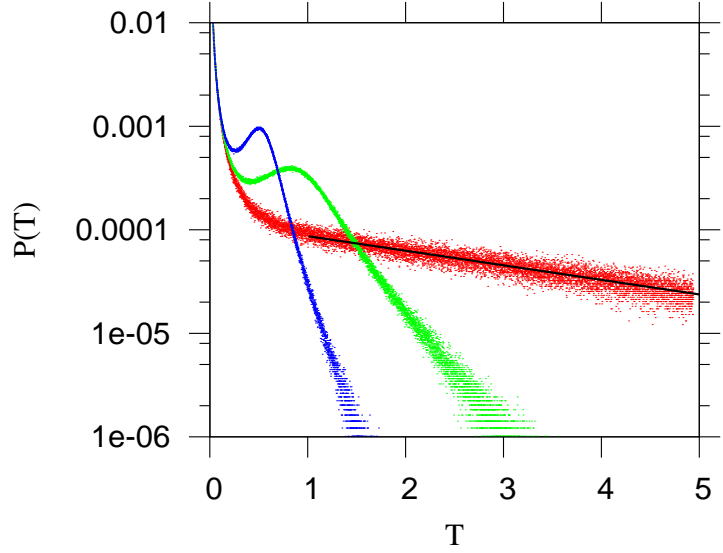}

\end{center}
\caption{\footnotesize (color online) Linear-log plot of $P(T)$ versus $T$. Values of the parameter $\alpha$ are respectively: $\alpha=0$ (the upper curve, with red dots), $\alpha=-0.1$ (the middle curve, with green dots) and $\alpha=-0.3$ (the lower curve, with blue dots). The fitted exponential function for the simple shear case ($\alpha=0$) is : $0.00012 \times \exp (-0.324 T)$ (where $\theta$ is used from method IV in section $3$). All data are for $N=10$ and $\dot{\gamma}=k=1$.
 }
\label{fig:fg2}
\end{figure}

Although IIA is inapplicable, we have simulated the Langevin dynamics of Eq.  (\ref{eq:langevin1}) and  curves for $P(T)$ were obtained (see Fig. \ref{fig:fg2}). The exponential tails fall more sharply ($\theta$ increases) with increasing negative $\alpha$. Furthermore the appearance of a peak (for $\alpha<0$), signifying a characteristic cyclic tumbling, gets more prominent with increasing negative $\alpha$.

\section{Special case of simple shear flow}
The limiting case of ``simple shear'' ($\alpha=0$) is the only
solvable case for $P(T)$ using the method of IIA \cite{das}, that too
for large shear rate $\text{Wi} \rightarrow \infty$. In the limit
$\alpha\rightarrow 0$, Eq.~\eref{eq:smallT-mixed} yields
\begin{equation}
A(T)= 1-\biggl[1+\frac{\pi^4\text{Wi}^2}{240}\biggr]^{-1}
\frac{4}{\pi^{3/2}} T^{1/2}
-\biggl[1+\frac{\pi^4\text{Wi}^2}{240}\biggr]^{-1}
\frac{\text{Wi}^2}{4} T^2 + \dotsb
\quad\text{as}~T\rightarrow 0.
\label{eq:smallT-shear}
\end{equation}
Now we note that in the limit of strong shear ($\text{Wi}\rightarrow
\infty$) the singular contribution of the second term proportional to
$\sqrt{T}$ in Eq. (\ref{eq:smallT-shear}) is ignorable. In the limit of $T\rightarrow 0$, 
Eq. (\ref{eq:smallT-shear}) has the limiting form 
$A(T)\approx 1-(60/\pi^4)T^2+\dotsb$ as $T\rightarrow 0$.  Hence, in
the limit $\text{Wi}\rightarrow \infty$ the process $R_x(t)$ becomes
``smooth'' \cite{majumder1} and IIA can be applied to find the
constant $\theta$. Clearly we have $A'(0)=0$ and
$A''(0)= -120/\pi$. Thus we get finite density of zero
crossings $\rho = \sqrt{120}/\pi^3$. If we assume $P(T)$ to be
exponential not only asymptotically but also over the full range of
$T$, the numerical value of $\rho$ gives a crude estimation of
$\theta$ as below:
\begin{equation}
\theta\approx \rho \approx 0.353298\cdots  \nonumber
\end{equation}


 We note that the expression of $A(T)$ for any $T$ in the limit of $\text{Wi}\rightarrow \infty$ is :

 \begin{eqnarray}
A(T) &=& \frac{960}{\pi^6}\sum_{p=1,3,5,...}^{\infty} e^{-p^{2}T}\left(\frac{1}{p^6}+\frac{T}{p^4}\right)
 \label{eq:A(T)-shear}
\end{eqnarray}

 Now we show below different estimates of $\theta$ from various IIA
 calculations.

{\bf(I)} First we consider the ``clipped'' variable $\chi(t)= \text{sgn} (R_x)$. Let $r(T)\equiv \frac{\langle \chi(t)\chi(t+\tau_0T)\rangle}{\langle \chi(t)^2\rangle}$ be the normalised asymptotic $(t\rightarrow \infty)$ correlator of $\chi(t)$.  One can express: $r(T)=\sum_{n=0}^{\infty}(-1)^np_n(T)$, where $p_n(T)$ is the probability that the interval $T$ has $n$ zeros. The strategy is to write the joint PDF of $n$ successive intervals as the product of the distribution of single intervals by IIA \cite{mcfadden,majumder3}. Then one ultimately finds a relation between the Laplace transforms of $P(T)$ and $ r(T)$  (which are denoted by $\tilde{P}(s)$ and $\tilde{r}(s)$ respectively) as below:
\begin{eqnarray}
\tilde{P}(s)=[1-(\rho/2)s(1-s\tilde{r}(s))]/[1+(\rho/2)s(1-s\tilde{r}(s))]
\label{eq:iiasatya}
\end{eqnarray}
where $\rho$ is the density of zero crossings. The asymptotic PDF  $P(T) \sim \exp(-\theta T)$ shows that $\tilde{P}(s)$ has a simple pole at $s=-\theta$ i.e the denominator in Eq. (\ref{eq:iiasatya}) has a simple zero at that point. Now, the correlator of the clipped process is given by: $r(T)=(2/\pi)\text{sin}^{-1}( A(T))$ (this relation holds for any Gaussian process \cite{rice}), where $A(T)$ is given by Eq. (\ref{eq:A(T)-shear}) . Using this relation in Eq. (\ref{eq:iiasatya}) and equating the denominator to zero we get (with a transformed variable $x=e^{-T}$):
\begin{eqnarray}
1+(\rho/2)s\left[1-\frac{2s}{\pi}\int_0^1x^{s-1}~\text{sin}^{-1}\left(\frac{960}{\pi^6}\sum_{p=1,3,5,...}^{\infty} x^{p^2}\left(\frac{1}{p^6}+\frac{\log {x}}{p^4}\right)\right) dx \right]= 0
\label{eq:iia}
\end{eqnarray}
Solving the above equation numerically for $s=-\theta$ we obtain :
\begin{equation}
 \theta \approx 0.356203\cdots \nonumber
\end{equation}

{\bf(II)} Two other IIA schemes are given in Ref. \cite{mcfadden}. Let $U(T)dT$ be the conditional probability that a zero of $\chi(t)$ occurs between $(t+T)$ and $(t+T+dT)$, given that there is a zero crossing of $\chi(t)$ at time $t$. It has been shown by J. A. McFadden that $\tilde{P}(s)$ and $\tilde{U}(s)$, the Laplace transforms of $P(T)$ and $U(T)$ respectively, are related by : $\tilde{P}(s)=\tilde{U}(s)/[1+\tilde{U}(s)]$. The expression of $U(T)$ for a Gaussian process is given by \cite{mcfadden} :

\begin{equation}
U=\frac{2}{\pi}\left(\frac{1}{H}+\text{tan}^{-1}H \right)\frac{r''}{4\rho}
\end{equation}
where 
\begin{equation}
r''=(2/\pi)M_{23}(1-A^2)^{-3/2}~~~~~~~;~~~~~~~H=M_{23}\left(M_{22}^2-M_{23}^2\right)^{-1/2} \nonumber
\end{equation}
\begin{equation}
M_{22}= - A ''(0)\left(1- A^2 \right)-A'^2~~~~~;~~~~~M_{23}= A ''\left(1-A^2 \right)+ A A'^2 \nonumber
\end{equation}

  The dependence on $T$ is understood in the above equations. We note that the function $\tilde{U}(s)$ raises some computational difficulty as it does not exist at $s=0$, because $U(T)$ does not vanish as $T \rightarrow \infty$. Also one can observe that as $T \rightarrow \infty$, $U(T)dT \rightarrow \rho dT$. If a new variable is defined as \cite{mcfadden}: $V(T)\equiv U(T)-\rho$, with Laplace transform $\tilde{V}(s)=\tilde{U}(s)-(\rho/s)$; then $\tilde{U}(s)$ can be substituted by $\tilde{V}(s)$ and the computation can be performed easily. Setting the denominator of $\tilde{P}(s)$ to zero we get the equation : $s+\rho+s\tilde{V}(s)=0$. Solving this equation numerically for $s=-\theta$ we have : 

\begin{equation}
\theta \approx 0.257136\cdots \nonumber
\end{equation}

{\bf(III)} In the third scheme, instead of assuming that all successive zero-crossing intervals are independent, it was assumed in Ref. \cite{mcfadden} that a given interval is independent of the sum of the $(2m+2)$ earlier intervals (for $m=0,1,2,...$). This is called ``quasi independent interval approximation''(QIIA). By QIIA one gets : $\tilde{P}(s)=(\tilde{U}(s)+g(s))/[2+\tilde{U}(s)-g(s)]$, where $g(s)=L\{r''(T)/4 \rho\}$. Again solving numerically for the zero of the denominator of $\tilde{P}(s)$ (using the newly defined variable $\tilde{V}(s)$ instead of $\tilde{U}(s)$) at $s=-\theta$ we get :
\begin{equation}
 \theta \approx 0.213558\cdots \nonumber
\end{equation}

{\bf(IV)} In Ref. \cite{das} another approximation $r(T) \approx A(T)$ was used that gives the value of $\theta$ very close to the numerics (see $\alpha=0$ case in Fig. \ref{fig:fg2}). This approximation yields :

\begin{equation}
\theta \approx 0.323558\cdots \nonumber
\end{equation}

 We see that various estimates of $\theta$ differ  slightly, but upto first decimal place (method III being an exception) the number is roughly $\approx 0.3$ .


\section{Conclusion}

We have exactly calculated the most general spatio-temporal correlation functions for a Rouse Polymer in planar mixed flow, in the steady state. Using the exact correlators of the mode coordinates, we explicitly write down the various correlators associated with the Cartesian components of the end-to-end vector $\mathbf R(t)$. The static correlators may be used to find the static probability distributions associated with the polar $(\vartheta)$ and azimuthal $(\phi)$ angles of $\mathbf R$. These static correlations give important information about the average shape of the polymer.

The more important concern in our paper has been the dynamics of angular tumbling of the vector $\mathbf R(t)$ in mixed flow, and the distribution of time scales associated with these first passage events. We find numerically that for rotationally dominated flows $(\alpha<0)$, the distribution has a peak and then an exponential tail for large times. We demonstrated that analytical estimates of the decay constant of the exponential tail can not be estimated using the method of IIA in $\alpha <0$ cases, as the process $R_x(T)$ is not smooth for any limiting value of $\text{Wi}$.

The method of IIA was successfully applied by us \cite{das} in the special case of ``simple shear'' $(\alpha=0)$, using a particular type of approximation (namely method IV in section $3$). Here we have given many new analytical estimates of the number $\theta$, using many different types of approximations following the classic work of Ref. \cite{mcfadden}. Although the numbers are slightly different from each other, up to the first decimal place $\theta \approx 0.3$.

\ack
We are thankful to RRI for the travel support which allowed the collaborators to meet and discuss this project. Dipjyoti Das  would also like to acknowledge Council of Scientific and Industrial Research (CSIR), India (JRF Award No. - 09/087(0572)/2009-EMR-I) for the financial support.

\appendix

\section{Static correlators }
\label{sec:AppA}
In this appendix, we give the exact expressions of the static correlators, which we get by putting $T=0$ in the equations (\ref{eq:correl1}) to (\ref{eq:correl6p}).

\underline{The static correlators for $\alpha>0$} :

 Using the standard series sums \cite{tables}:
\begin{equation}
\sum_{m=1}^{\infty}\frac{1}{(2m-1)^2-z^2}=\frac{\pi}{4z}\tan \frac{\pi z}{2}~~~;~~~\sum_{m=1}^{\infty}\frac{1}{(2m-1)^2+z^2}=\frac{\pi}{4z}\tanh \frac{\pi z}{2} ~~,
\end{equation}

we simplify 

\begin{equation}
d=\langle R_x^2 \rangle  =  \frac{2\zeta\tau_0 }{N} \left[ (1+1/\alpha) \sum_{p=1,3,5,...}^{\infty} \left(\frac{p^2}{p^4-\text{Wi}^2\alpha}\right)+(1-1/\alpha)\frac{\pi^2}{8}\right]
\end{equation}

to get :

\begin{equation}
d=\langle R_x^2 \rangle  = \frac{2\zeta\tau_0 }{N}\left[(1+1/\alpha)\frac{\pi}{8x}\left(\tan \frac{\pi x}{2}+\tanh \frac{\pi x}{2}\right)+(1-1/\alpha)\frac{\pi^2}{8}\right]
\end{equation}

Similar steps yield the following :
\begin{equation}
a=\langle R_y^2 \rangle  = \frac{2\zeta\tau_0 }{N}\left[(1+\alpha)\frac{\pi}{8x}\left(\tan \frac{\pi x}{2}+\tanh \frac{\pi x}{2}\right)+(1-\alpha)\frac{\pi^2}{8}\right]
\end{equation}
\begin{equation}
b=\langle R_xR_y \rangle  = \frac{2\zeta\tau_0 }{N}\left[\frac{(1+\alpha)}{\sqrt{\alpha}}\frac{\pi}{8x}\left(\tan \frac{\pi x}{2}-\tanh \frac{\pi x}{2}\right)\right]
\end{equation}
\begin{equation}
c=\langle R_z^2 \rangle  = \frac{\zeta\tau_0\pi^2 }{2N}
\end{equation}

where $x=\sqrt{\text{Wi}}\alpha^{1/4}$.

\underline{The static correlators for $\alpha<0$} :
\begin{equation}
\bar{d}=\langle R_x^2 \rangle  = \frac{2\zeta\tau_0 }{N}\left[(1+1/\alpha)\frac{\pi}{8y}\frac{\left(\tan \frac{\pi y}{2}\text{sech}^2 \frac{\pi y}{2}+\tanh \frac{\pi y}{2}\text{sec}^2 \frac{\pi y}{2} \right)}{(1+\tan^2 \frac{\pi y}{2}\tanh^2 \frac{\pi y}{2})}+(1-1/\alpha)\frac{\pi^2}{8}\right]
\end{equation}
\begin{equation}
\bar{a}=\langle R_y^2 \rangle  = \frac{2\zeta\tau_0 }{N}\left[(1+\alpha)\frac{\pi}{8y}\frac{\left(\tan \frac{\pi y}{2}\text{sech}^2 \frac{\pi y}{2}+\tanh \frac{\pi y}{2}\text{sec}^2 \frac{\pi y}{2} \right)}{(1+\tan^2 \frac{\pi y}{2}\tanh^2 \frac{\pi y}{2})}+(1-\alpha)\frac{\pi^2}{8}\right]
\end{equation}

\begin{equation}
\bar{b}=\langle R_xR_y \rangle  = \frac{2\zeta\tau_0 }{N}\left[\frac{(1+\alpha)}{\sqrt{|\alpha|}}\frac{\pi}{8y}\frac{\left(\tan \frac{\pi y}{2}\text{sech}^2 \frac{\pi y}{2}-\tanh \frac{\pi y}{2}\text{sec}^2 \frac{\pi y}{2} \right)}{(1+\tan^2 \frac{\pi y}{2}\tanh^2 \frac{\pi y}{2})}\right]
\end{equation}
\begin{equation}
\bar{c}=\langle R_z^2 \rangle  = \frac{\zeta\tau_0\pi^2 }{2N}
\end{equation}
where $y=\sqrt{\text{Wi}/2}|\alpha|^{1/4}$. We checked that in the limit $\alpha \rightarrow 0+$ as well as $\alpha \rightarrow 0-$ the above static correlators smoothly go to the exact expressions of the static correlators in ``simple shear'' that are published in Ref. \cite{das}.



\end{document}